\documentclass[12pt,preprint]{aastex}

\newcommand{\error}[2]{\mbox{$#1 \pm #2$}}
\newcommand{\errorBoth}[3]{\mbox{$#1^{+#2}_{-#3}$}}
\newcommand{\north}{pos-NN}

\newcommand{\eastright}{pos-ER}
\newcommand{\eastup}{pos-EU}

\shortauthors{Miyata et al.} 
\shorttitle{Reflection Shocked Gas in the Cygnus Loop} 

\begin{document}

\title{Reflection Shocked Gas in the Cygnus Loop Supernova Remnant}
\author{Emi Miyata\altaffilmark{1} and Hiroshi Tsunemi\altaffilmark{1}}
\affil{Department of Earth and Space Science,
	Graduate School of Science, Osaka University \\
	 1-1, Machikaneyama, Toyonaka, Osaka, 560-0043, Japan}
\email{miyata@ess.sci.osaka-u.ac.jp, tsunemi@ess.sci.osaka-u.ac.jp}

\altaffiltext{1}{CREST, Japan Science and Technology}

\begin{abstract}
 We performed spectroscopic X-ray observations of the eastern and
 northern regions of the Cygnus Loop with the ASCA observatory. The
 X-ray surface brightness of these regions shows a complex structure in
 the ROSAT all-sky survey image. We carried out a spatially-resolved
 analysis for both regions and found that $kT_{\rm e}$ did not increase
 toward the center region, but showed inhomogeneous structures. Such
 variation cannot be explained by a blast wave model propagating into a
 homogeneous interstellar medium. We thus investigated the interaction
 between a blast wave and an interstellar cloud. Two major emission
 mechanisms are plausible: a cloud evaporation model and a reflection
 shock model. In both regions, only a reflection shock model
 qualitatively explains our results.  Our results suggest the existence
 of a large-scale interstellar cloud. We suppose that such a large-scale
 structure would be produced by a precursor.
\end{abstract}

\keywords{ISM: individual (Cygnus Loop) -- supernova remnants
		--- X-rays: Spectra}

\pagebreak

 \section{INTRODUCTION}

 The Cygnus Loop is the prototype of evolved shell-like supernova
 remnants (SNR). Its large apparent size and high surface brightness make
 it an ideal object for the study of the interaction between the
 supernova blast wave and the interstellar matter or the interstellar
 cloud in detail. Theoretical calculations of these interactions have
 been carried out by many authors (Sgro 1975; McKee \& Cowie 1975; McKee
 \& Ostriker 1977; Spitzer 1982; McKee 1987; Klein, McKee \& Colella
 1994).

 Various optical and ultraviolet observations of the Cygnus Loop have
 been performed to study the shock waves and their interactions. Cox
 (1972) suggested that the Cygnus Loop is in the radiative cooling phase
 of its evolution. X-ray observations, however, revealed that the Cygnus
 Loop is in the adiabatic phase and is still evolving (e.g. Ku et
 al.1984; Charles et al. 1985; Miyata et al. 1994). McKee \& Cowie
 (1975) modeled the interaction between a blast wave and an interstellar
 cloud and were able to explain the discrepancies between the optically 
 measured velocities of the filaments and the blast wave velocity
 inferred from X-ray observations.

 There are two models to explain the apparent optical and X-ray
 emission; an evaporation model and a reflection shock model. The
 evaporation model was proposed by McKee \& Cowie (1975). They modeled a
 blast wave propagating into an inhomogeneous interstellar medium where
 there are many small clouds (cloudlets). In this model, the cloudlets
 are not directly heated by the shock wave but are evaporated due to
 thermal conduction by the ambient hot gas. This model was applied to
 both optical data (Fesen, Blair, \& Kirshner 1982) and X-ray data
 (Tsunemi \& Inoue 1980; Ku et al. 1984). It was, however, ruled out for
 the eastern region of the Cygnus Loop (e.g. Hester \& Cox 1986,
 hereafter HC).  Recent observations of [\ion{Ne}{5}] emission at the
 eastern region also suggests that thermal evaporation does not play an
 important role (Szentgyorgyi et al. 2000).

 The reflection shock is important when the cloud size is large ($\sim$
 1pc). Recent optical and X-ray data have been explained in the context
 of this model at various portions of the Cygnus Loop; the southeastern
 blob (SE blob) by Graham et al. 1995, the eastern region by HC and
 Raymond et al. 1988, and the northeastern (NE) region by Hester,
 Raymond \& Blair 1994.

 Such a large scale cloud reminds us of a pre-existing cavity wall
 produced by the precursor, as was first proposed by McCray \& Snow
 (1979).  Shull \& Hippelein (1991) measured proper motions at 39
 locations within the Cygnus Loop and found an asymmetric expansion of
 the shell. They also concluded the existence of a pre-existing cavity
 wall around the Cygnus Loop. This model is also supported by
 observations of coronal iron line emission (Teske 1990), observations
 with IRAS (Braun \& Strom 1986) and observations with the Rosat HRI
 (Levenson et al. 1997).  Miyata \& Tsunemi (1999, hereafter MT) found a
 large density discontinuity at the NE region. Combined with a large
 estimated mass of the progenitor star, they also supposed the existence
 of a pre-existing cavity.

 In this paper, we carried out spatially-resolved X-ray
 spectroscopy of the eastern and  northern regions of the Cygnus
 Loop. Data analyzed here were obtained with the ASCA Observatory and
 the ROSAT Observatory.

 \section{OBSERVATIONS AND DATA CORRECTIONS}

 We observed the eastern and northern regions of the Cygnus Loop in
 1993-1994 with the ASCA observatory (Tanaka, Inoue, \& Holt 1984).
 Figure~\ref{fig:ana:limb:image} shows the location of SIS FOVs with
 black squares superimposed on an X-ray surface brightness map of the
 Cygnus Loop (Aschenbach \& Leahy 1994).  A detailed observational log
 is given in Table~\ref{obslog}.  We used the Faint-mode data obtained
 with the SIS to achieve the best spectral resolution (Yamashita et
 al. 1997).  We excluded the data taken at an elevation angle below
 5$^\circ$ from the night earth rim and 50$^\circ$ from the day earth
 rim, a geomagnetic cutoff rigidity lower than 6 GeV c$^{-1}$, and the
 region of the South Atlantic Anomaly. We also eliminated times when the
 pulse heights of corner pixels rapidly change.  We then removed the hot
 and flickering pixels and corrected CTI, DFE, and Echo effects (Dotani
 et al. 1995; Dotani et al. 1997) in our data sets by using {\tt sispi}
 and {\tt faint} commands. Since we performed the observation of the
 northern region in the ASCA PV phase, the residual dark distribution
 (RDD) effect is not serious even for 4CCD mode data. However, the data
 of the eastern region observed in 1994 were degraded due to the RDD
 effect (see detail in Dotani 1995). Therefore, we corrected the RDD
 effect for the eastern region using the frame data obtained near to the
 epoch of our observation. We retrieved the frame data taken for the
 calibration observation of Cyg X-1 and anonymous sky (their sequence
 numbers are 18941107, 10021040 and 10021060) since they were observed
 in 1994 November.

 X-ray spectra accumulated in four regions are shown in
 Figure~\ref{all_spec}. We can confirm that the X-ray emission is
 dominated by thermal emission and emission lines from Mg and Si are
 clearly detected whereas emission lines from O, Ne, and Fe are weak.

 \section{DATA ANALYSIS}

 We investigated the spatial distribution of physical parameters using
 the SIS data in detail, using the same method as in MT in the northern
 region.  In the northern region, we divided our FOV into 3\arcmin\
 squares each of which corresponds to the half-power diameter of the XRT
 (Serlemitsos et al. 1995).  Furthermore, we arranged the small squares
 such that each square overlapped adjacent squares by 2\arcmin\,
 resulting in 400 spectra.

 In the eastern region, however, the surface brightness is lower than
 that of the northern region as shown in Table~\ref{obslog}. We divided
 our FOV into 5\arcmin\ in order to obtain statistics similar to 
 those for the northern region. Each small region is arranged such that
 each square overlapped adjacent squares by 2\arcmin\,
 resulting in 49 spectra from each eastern region.

 Because the optical axes of SIS0 and SIS1 were different from each other,
 we employed the {\tt ascatool} library to adjust the optical axes.
 After adding the spectra of SIS-0 and SIS-1, we applied the single
 component non-equilibrium ionization (NEI) model coded by K. Masai
 (Masai 1984; Masai 1994) to each spectrum. Free parameters of the
 fitting were $kT_{\rm e}$, ionization parameter (log($\tau$)),
 abundances of O, Ne, Mg, Si, Fe, and Ni, and emission measure (EM
 in cm$^{-6}$ pc).  Values for the reduced $\chi^2$ ranged from 1 to 2.

  \subsection{Plasma Structure of the Eastern Region}

 Figure~\ref{fig:ana:east:kt} shows the spatial variation of $kT_{\rm
 e}$ and log($\tau$) for the eastern region.  We find a complicated
 $kT_{\rm e}$ structure revealed by the SIS. The values of $kT_{\rm e}$
 do not smoothly increase toward the center of the Cygnus Loop, but have
 local maxima at the crossing point of three observations ($\sim$ 0.6
 keV) and the northern region of \eastup\ ($\sim$ 0.8 keV). This
 $kT_{\rm e}$ distribution cannot be explained with a blast wave model
 propagating into a homogeneous medium. On the contrary, the spatial
 variation of log$(\tau)$ is relatively uniform with $10.3 \pm 0.4$.  We
 can confirm the hot plasma is still in the ionizing phase.

   \subsubsection{Comparison with the ROSAT Image}

   We retrieved the ROSAT PSPC data from the HEASARC/GSFC. The sequence
   number was RP500090.  The X-ray surface brightness map obtained with
   the ROSAT PSPC is shown in Figure~\ref{fig:ana:east:pspc} (upper).
   Black squares show the FOVs of the SISs. Vignetting corrections were
   performed using the {\it EXSAS/MIDAS} software package.  We see the
   instrumental structure due to the ribs of the window support of the
   PSPC which are seen only in a direction parallel to the direction of
   the telescope wobble. We found significant inhomogeneities in our
   FOVs. They could not be explained with a simple blast wave model.
   There is a bright clump (c1) at the PSPC center ($\alpha = 20^{\rm h}
   57^{\rm m} 14^{\rm s}, \delta = 31^\circ 1^\prime 55^{\prime\prime}$)
   with a radius of 2.$^{\!\prime}$7 ($\sim$ 0.6~pc) where we guess that
   the blast wave is moving into a dense region and has stagnated. The
   region named c1 is the bright part of ``XA'' named in HC.  The
   regions to the north and south of c1 seem to have lower density,
   which means that the blast wave goes ahead of c1. There is another
   bright clump to the south of c1 which is the SE blob.  To the west of
   c1, we see a bright arc structure which is the left portion of
   \eastright.

   The lower panel of Figure~\ref{fig:ana:east:pspc} shows the surface
   brightness map obtained with the SIS. Comparing these two images, we
   notice that c1, marked with a black circle, is not bright in the SIS
   image. The bright regions in the ASCA energy band (0.5$-$10 keV) are
   the northeast region of \eastright\ and the southwestern region of
   \eastup. The effective energy range of the PSPC is 0.1$-$2.4 keV
   while that of the SIS is 0.5$-$10 keV.  There is a big difference in
   the sensitivity below the C-K edge.  Therefore, possible
   interpretations of the discrepancies between the ROSAT and ASCA
   images are 1) the presence of a low $kT_{\rm e}$ component at c1, 2)
   the presence of a high $kT_{\rm e}$ component at ASCA bright region,
   and/or 3) an extremely low $N_{\rm H}$ around c1.

   \subsubsection{Band Ratio Map of the PSPC}

   We constructed a band ratio map using the PSPC data by dividing the
   0.44$-$2.0 keV band (R3$-$R7) image by the 0.11$-$0.28 keV band
   (R1$-$R2) image, shown in Figure~\ref{fig:ana:east:hardness}.  We
   clearly see a soft X-ray emitting region along the shock front in the
   FOV.  The softest region is at c1 in this figure, shown in white. We
   also found a soft X-ray emitting region around c1 as shown by a
   circle (hereafter referred to as c2). Around c2, a hard X-ray
   emitting region extends and wraps around c2 where $kT_{\rm e}$
   determined with the SIS is also high as shown in
   Figure~\ref{fig:ana:east:kt}.
   
   \subsubsection{Combined Spectral Analysis}

   As shown in Figure~\ref{fig:ana:east:hardness}, a spectral variation
   is seen in the PSPC data.  We analyzed the PSPC spectra at various
   positions within the FOV.  We used {\tt xselect} to extract the X-ray
   spectra, and used the response function { \tt pspcb\_gain2\_256.rmf}
   for the spectral fits. The effective areas of the XRT were calculated
   by {\tt pcarf}, which included vignetting
   effects. Figure~\ref{fig:ana:east:hardness} shows the five regions
   (c1, r1$-$r4) where spectra were extracted.  The spectrum obtained for
   each region is shown in Figure~\ref{fig:ana:east:pspcSpec}.  The
   background spectrum was extracted from a blank region in the same
   data set.  We clearly see strong soft X-ray emission below the C
   K-edge in all regions.  This soft excess could account for the
   difference in the surface brightness maps between the ROSAT and ASCA
   energy bands.

   We extracted the SIS spectra from the same region as the PSPC (c1,
   r1$-$r4) and simultaneously fitted the PSPC and SIS spectra.  We
   applied the NEI model for these spectra. All parameters except the
   normalization were linked between the SIS and PSPC spectra. Free
   parameters were $kT_{\rm e}$, log($\tau$), abundances of O, Ne, Mg,
   Si, and Fe, $N_{\rm H}$, normalizations for both the SIS spectra and
   the PSPC spectra. We could not obtain statistically acceptable fits
   with a single component NEI model. In all PSPC spectra, we found
   excess emission below the C K-edge. We confirmed that such excess
   emission could not be reproduced with extremely low $N_{\rm H}$ such
   as $10^{19}$ cm$^{-2}$.  We therefore added another component with a
   different $kT_{\rm e}$ to reproduce the low energy region of the
   spectra. The low $kT_{\rm e}$ component was mainly detected in the
   PSPC energy band. Due to the insufficient energy resolving power of
   the PSPC, we could not determine the individual abundance of heavy
   elements and log($\tau$) for the low $kT_{\rm e}$ component. We
   therefore assumed cosmic abundances and the condition of ionization
   equilibrium for the low $kT_{\rm e}$ component. We obtained
   acceptable fits for all regions.  Best fit curves are shown in
   Figure~\ref{fig:ana:east:pspcSpec} and parameters are summarized in
   Table \ref{east_fit_result} and \ref{east_abundance}.

   Including the ROSAT PSPC (C-band), we found a very low $kT_{\rm e}$
   component. The ASCA SIS is insensitive to such a low $kT_{\rm e}$
   component.  The $kT_{\rm e}$ values obtained for high $kT_{\rm e}$
   components are similar in all regions and also similar to those
   obtained at the NE region (MT). The values of log($\tau$)
   are also similar in all regions whereas they are slightly higher than
   those in the NE region.

   Since the $kT_{\rm e}$ value obtained for low $kT_{\rm e}$ component
   is extremely low, the major part of its emission is below the PSPC
   band.  Therefore, the $kT_{\rm e}$ value obtained might contain some
   uncertainties because we only observe a small fraction of the
   spectrum.  Moreover, the emissivity of such low-$kT_{\rm e}$ plasma
   is very large (typically $10^{-22}$ erg s$^{-1}$ cm$^{-3}$ for plasma
   with cosmic abundances; Gehrels \& Williams 1993), resulting to
   quickly cool down to $\simeq 10^5$ K.  Other narrow band studies,
   however, reveal the existence of such low $kT_{\rm e}$ plasma.
   Images of [\ion{Fe}{10}] and [\ion{Fe}{14}] lines showed the
   existence of the cooler X-ray emitting plasma in the vicinity of the
   eastern region of the Cygnus Loop (Teske 1990; Teske \& Kirshner
   1985; Ballet et al. 1984).  The image of the [\ion{Ne}{5}] line
   showed bright filaments in the vicinity of c1, r1, and r2 regions
   (Szentgyorgyi et al. 2000) whereas it did not cover other regions.
   As [\ion{Ne}{5}] is a unique probe of plasma with $(1-6)\times 10^5$
   K, we confirm that the extremely low $kT_{\rm e}$ component is surely
   present at c1 and r1.

   We should note that the metal abundances are depleted at c1 and r4
   while they are comparable to the cosmic values at r1, r2 and r3.
   This reminds us the spatial structure detected at the NE region of
   the Cygnus Loop.  In the NE region, abundances of heavy elements are
   also depleted and shows an increase toward the inner region (MT).  At
   the shell, a radius of $\geq 0.9 R_s$ in the NE region where $R_s$ is
   the shock radius, they are constant (0.15$\sim$0.3 times cosmic
   values), whereas, behind the shell, they are 20$-$30\% larger than
   those obtained at the shell.  Therefore, abundances at c1 and r4 are
   consistent with those obtained at the shell whereas those obtained at
   r1, r2, and r3 are qualitatively consistent with those obtained
   behind the shell with considering the statistical errors.

   \subsection{Plasma Structure of the Northern Region}\label{sec:ana:north}

   Figure \ref{fig:dis:north:kt} shows the spatial variations of
   $kT_{\rm e}$ and log($\tau$) in the \north.  As shown in
   Figure 1, a V-shaped structure is clearly seen over the SIS
   FOV. Values of $kT_{\rm e}$ at the V-shaped structure ($\simeq$ 0.3
   keV) are lower than those in the neighboring region. This is
   consistent with results obtained with the ROSAT PSPC (Decourchelle et
   al. 1997). There are three high $kT_{\rm e}$ regions marked in Figure
   ~\ref{fig:dis:north:kt}. Such inhomogeneities in the $kT_{\rm e}$
   distribution cannot be explained by a simple blast wave model. The
   spatial variation of log($\tau$) shows an anti-correlation with that
   of $kT_{\rm e}$.

   \subsection{Comparison with the FPCS Results}

   One of the most important purposes of observing the northern region is
   to compare the results with those of the FPCS (Vedder et al. 1986).
   The FPCS onboard the Einstein Observatory was the
   only available data with better energy resolution than ours and had a
   capability to resolve the triplet of \ion{O}{7}; forbidden (561 eV),
   inter-combination (569 eV), and resonance (574 eV) lines.

   To compare our results with those of the FPCS, we investigated the
   intensity of individual emission lines. We fitted the spectrum,
   extracted from the same region as Vedder et al. (1986), with a
   thermal bremsstrahlung continuum and then added a Gaussian line
   profile one by one until an acceptable fit was obtained (combination
   model as mentioned in Miyata et al 1998).
   Figure~\ref{fig:ana:sis:north:bg} shows the best fit curve and the
   residuals between the data and the model. An acceptable fit was
   obtained with a model that included 10 Gaussian line profiles.  The
   best fit $kT_{\rm e}$ and line intensity ratio of \ion{O}{8} to
   \ion{O}{7} values are \error{0.31}{0.01} keV and \error{0.91}{0.06}.

   Due to its insufficient energy resolution, the SIS cannot resolve the
   \ion{O}{8} K$\alpha$ line from the \ion{O}{7} K$\beta$ line. We then
   calculated the line intensity ratio of the sum of \ion{O}{7} K$\beta$
   and \ion{O}{8} to \ion{O}{7} based on the Table~1 in Vedder et
   al. (1986). Thus, the values obtained with the FPCS are $kT_{\rm e}
   \geq$ 0.3 keV and a line ratio of \error{1.5}{0.6}, respectively.
   The SIS results fully agree with the FPCS results to within the
   statistical uncertainties.  We therefore confirm the consistency
   between the SIS and the FPCS.

   \section{INTERACTION BETWEEN THE BLAST WAVE AND INTERSTELLAR CLOUD}

   As mentioned in section 3, a simple blast wave model is able to
   explain the inhomogeneities neither in the eastern nor in the
   northern regions. We need to introduce interstellar clouds into the
   model in order to account for the inhomogeneities. The interaction of
   the shock wave with the interstellar clouds results in three kinds of
   shock; a shock traveling around the cloud, a shock transmitted into
   the cloud (hereafter referred to as the transmission shock), and a
   shock reflected from the cloud (hereafter referred to as the
   reflection shock).  The nature of the interaction depends on the size
   and the density of the cloud.

   \subsection{Reflection Shock Model}\label{sec:review:reflection}

   After the interaction, the reflection shock will propagate toward a
   direction opposite to the incident shock when the postincident shock
   flow is supersonic relative to the cloud.  We assume radiation can be
   ignored and that the cloud is spherical.  If the density contrast
   between the cloud and the ambient matter is not high enough to be
   treated as incompressive, some amount of energy will be released by
   the transmission shock, propagating into the cloud.

   The physical properties of the transmission shock as well as the
   reflection shock depend on the density contrast, $\alpha = n _4 / n
   _0$, where $n _4$ is the density of the cloud. Sgro (1975) studied
   the interaction by considering the density contrast in the assumption
   of the strong shock limit for the blast wave (Mach number ($M_0) \
   \geq$ 6). He solved the fluid equations for the mass, momentum, and
   energy conservation across the discontinuity between the reflection
   shock and the transmission shock. The geometry of the interaction is
   shown in Figure~\ref{fig:review:reflection:geometry}.  We denote by
   subscripts 0, 1, 2, 3, and 4 the parameters of preshocked gas,
   postincident shocked gas, postreflected shocked gas, posttransmitted
   shocked gas, and incident cloud, respectively, as shown in this
   figure.  Across the discontinuity, the pressure is constant, $p_2 =
   p_3$.  When the postincident shock wave strikes the cloud, it would
   experience an overpressure relative to the ambient postincident
   shocked gas by a factor of $\beta$.
   \begin{equation}\label{eq:review:reflection:preEq}
    n _3 v_3^2 = \beta n _1 v_1^2
   \end{equation}
   He found the following equations for temperature and velocity, with
   respect to the contact discontinuity, of both the reflection and
   transmission shock region as a function of postincident shocked
   parameters.
   \begin{eqnarray}\label{eq:review:reflection:param}
    T_3 &=& \frac{\beta}{\alpha} \ T_1 \\
    T_2 &=& \frac{\beta}{\alpha _r} \ T_1 \\
    v_3 &=& \sqrt{\frac{\beta}{\alpha}} \ v_1 \\
    v_2 &=& \left\{\frac{3}{4} - \frac{1}{4}
    \left(\frac{15 \alpha_{r}}{4-\alpha_{r}}\right)^{1/2}\right\}
    v_1\ \ \ , \\
   \end{eqnarray}
   where $\alpha _r$ is the density contrast between the postincident shocked
   gas and the postreflected shocked gas, $n _2 / n _1$, and can
   be related to $\alpha$ through
   \begin{equation}
    \alpha  = \frac{3 \alpha _r \left(4\alpha _r - 1\right)}
     {[\{3 \alpha _r \left(4-\alpha _r\right)\}^{1/2} - \sqrt{5}
     \left(\alpha _r - 1\right)]^2}
     \ \ \ .
   \end{equation}
   Figure~\ref{fig:dis:east:sgro} shows the characteristics of the
   postreflected shocked region as a function of $\alpha$.  This figure
   is consistent with Figure 17 of Hester et al. (1994) even though
   their approach of calculation differs from ours.

   \subsection{Cloud Evaporation Model}\label{sec:review:evaporation}

    When the postincident shock wave strikes the cloud, it would
    experience an overpressure relative to the ambient postincident
    shocked gas by a factor of $\beta$. Values of temperature and
    velocity for the transmission shock can be determined by equations
    (\ref{eq:review:reflection:preEq},\ref{eq:review:reflection:param}).
    The time required for the transmission shock to travel through the
    cloud is $10^6 R_4 / v_3 $ yr, where $R_4$ is the radius of the
    cloud in pc and $v_3$ is the velocity of the transmission shock in km
    s$^{-1}$.  By equating the cooling time of the posttransmitted
    shocked gas to the time required for the transmission shock to
    travel through the cloud, a critical cloud density, $n_c$, is
    defined. Sgro (1975) calculated $n_c$ and the cooling time, $\tau
    _c$, of the posttransmitted shocked gas due to the radiation as
   \begin{equation}\label{eq:review:evapo:criticalDensity}
    { n_c \simeq 10
     \left(\frac{\beta}{3.2}\right)^{5/7}
     \left(\frac{R_4}{1\ \rm pc}\right)^{-2/7}
     \left(\frac{n_0}{0.5\ {\rm cm}^{-3}}\right)^{5/7}
     \left(\frac{T_1}{10^7 \ \rm K}\right)^{5/7} \\
    \,\,\, {\rm cm}^{-3},}
   \end{equation}
   \begin{equation}
    \tau_c = 4.6\times 10^6 \left(\frac{n_4}{n_0}\right)^{-3}
     \beta^2 v_1^4 \ \ \ \ {\rm yr},
   \end{equation}
   where $\epsilon$ is the emissivity of thin thermal gas, which is
   approximately 1.33$\times 10^{-16} T^{-1}$
   ergs cm$^{-3}$ s$^{-1}$ in the range of
   $10^5 \leq T \leq 10^7$ K.

   For clouds with density greater than $n_c$, the
   posttransmitted shocked gas cools rapidly and clouds begin to evaporate
   into a hot tenuous medium. Note that $\beta$ depends on the density 
   contrast, $\alpha$ = $n_c$/$n_0$, so that we should pay attention to
   calculate $n_c$.
   
   Figure~\ref{fig:dis:east:nc} shows the maximum cloud radius which can
   be heated up by the blast wave.  $\beta$ is calculated for each
   density contrast and $T_1$ is fixed to be 0.21 keV. This graph
   allows us to determine $n_c$ for a given $n_0$ and $R_c$.
   
   After the cloud is surrounded by the tenuous medium, it will be
   gradually heated up due to thermal conduction. The thermal
   conductivity of a fully ionized Hydrogen plasma is given by (Spitzer
   1962)
   \begin{equation}
    \kappa = 1.94 \times 10^{11} \left(\frac{T_e}{10^7 \ \rm K}\right)^{5/2}
     \left(\frac{\ln\Lambda}{30}\right)^{-1}
     \ \ \ {\rm ergs\ s^{-1} K^{-1} cm^{-1}} \ \ \ ,
   \end{equation}
   where $\ln\Lambda$ is the Coulomb logarithm which we set to
   $\ln\Lambda$=30 since it is insensitive to $T_{\rm e}$ and density.
   We set the differences in temperature between the cloud and the
   surrounding medium to be $\Delta T$, so that the heat flux is
   \begin{equation}
    q = - \kappa \Delta T\ \ \ .
   \end{equation}
   The heat flux is proportional to $\lambda$/$L_T$, where
   $\lambda$ is the mean free path for electron energy exchange and
   $L_T$ is the temperature scale height $T/| \Delta T |$.
   $\lambda$ is found from 
   \begin{eqnarray}\label{eq:review:mfp}
    \lambda &=&  t_{\rm ee} \left(\frac{3 kT_{\rm e}}{m_{\rm e}}\right)^{1/2} \nonumber\\
    &=& 0.29 \ \left(\frac{T_e}{10^7 \ \rm K}\right)^2
     \left(\frac{n_e}{1 \rm cm^{-3}}\right)^{-1}
     \left(\frac{\ln\Lambda}{30}\right)^{-1}
     \ \ \ {\rm pc} \ \ \ .
   \end{eqnarray}
   
   The classical model of thermal conductivity assumes that the mean free
   path, $\lambda$, is shorter than $L_T$, which is a typical cloud
   size.  However, for many situations of interest to our model, the mean 
   free path of electrons is not shorter than the radius of the evaporating
   cloud.  When $\lambda$ becomes comparable to or greater than $L_T$,
   the heat flux is no longer equal to $-\kappa \Delta \rm T$. This
   effect is referred to as saturation.  Cowie \& McKee (1977) studied 
   `saturated' thermal conduction in detail.  As to the spherical cloud,
   the timescale for complete evaporation can be written by (Cowie \& McKee
   1977)
   \begin{equation}\label{eq:review:evapo:timescale}
     t_{evp} \simeq 1.8\times 10^4 
     \left(\frac{n_4}{17}\right)
     \left(\frac{R_4}{1 \ \rm pc}\right)^2
     \left(\frac{T_1}{10^7 \ \rm K}\right)^{-5/2}
     \left(\frac{\ln\Lambda}{30}\right)
     \ \ \ {\rm yr}\ \ .
   \end{equation}

  \section{SHOCK STRUCTURE OF THE EASTERN REGION}

   \subsection{Density of Bright Clump}

   As shown in Figure~\ref{fig:ana:east:hardness}, we found two soft
   X-ray bright regions in the eastern region; c1 and c2.  Inside c2, we
   detected low $kT_{\rm e}$ components at every position (c1 and
   r1$-$r4). Based on these results, we calculated the density of these
   regions. We assumed a homogeneous spherical cloud for c2 with a
   radius of 12\arcmin\ $\approx$ 2.7~pc for the low $kT_{\rm e}$
   components.  The calculated volumes and densities are summarized in
   Table~\ref{table:dis:east:pspcSpec}.  If we assumed pressure balance
   between the low $kT_{\rm e}$ and high $kT_{\rm e}$ components, we
   could calculate the density and volume for the high $kT_{\rm e}$
   components as shown in Table \ref{table:dis:east:pspcSpec}.

   The obtained density for the low $kT_{\rm e}$ component is roughly 30
   cm$^{-3}$ at every portion of the eastern region. The density for the high
   $kT_{\rm e}$ component is roughly 5 times higher than that obtained
   at the NE region as found in MT.

   We also calculated the thermal pressure at each region shown in Table
   \ref{table:dis:east:pspcSpec}. HC estimated the
   thermal pressure at the ``XA''to be $(1.1-3.8) \times 10^7$ cm$^{-3}$
   K, which is consistent with our results.

   \subsection{Emission Mechanism of the Eastern Region}\label{sec:dis:east:em}

   Because a simple blast wave model can not explain the inhomogeneities
   of the eastern region, we investigate emission mechanism models which can
   hold for the overall X-ray features of the eastern region.

    \subsubsection{Locally Large Preshock Density}

   We first consider an inhomogeneous ISM. Specifically, we consider an
   enhancement of the preshock density. To distinguish it from the
   interstellar cloud, we assume that the enhanced density is lower than
   $n_c$.  In this case, such an enhancement will be smoothed out by the
   blast wave, resulting in an enhancement of the X-ray emission. This
   could account for the bright X-ray appearance at the eastern region.

   Based on this model, $kT_{\rm e}$ of the bright region is expected to be
   lower than those of neighboring regions. The high $kT_{\rm e}$ region can
   be considered as the postincident shocked region.  This means that we
   can apply a simple blast wave model to the high $kT_{\rm e}$ region
   as a first order approximation.  Based on the Sedov model (Sedov
   1959), $kT_{\rm e}$ just behind the blast wave is proportional to
   $n^{-2/5}$.  For the high $kT_{\rm e}$ components, $n_{\rm e}$ is roughly
   6 cm$^{-3}$. Assuming the strong shock limit, $n_{\rm 0}$ is calculated
   to be 1.5 cm$^{-3}$. This value is roughly 6 times higher than that
   of the NE region (MT). Considering the $kT_{\rm e}$ value at the NE
   region, $kT_{\rm e}$ at the eastern region should be $(1.5/0.25)^{-2/5}
   \times 0.3\ \approx$ 0.15~keV. This value is roughly two times lower
   than that obtained at c2 region as shown in
   Table~\ref{table:dis:east:pspcSpec}. Therefore, the $kT_{\rm e}$ value
   obtained for the high $kT_{\rm e}$ component is too high to be explained 
   using the simple blast wave model.

   Thus, it seems difficult to explain the current bright X-ray features
   as resulting from a shocked region in which the density is higher than 
   that of the neighboring region.

    \subsubsection{Cloud Evaporation Model} 

    We found region c2 to be bright in soft X-rays. As shown
    in Table~\ref{table:dis:east:pspcSpec}, the density of this region
    is quite high compared with other regions (0.25 cm$^{-3}$ at the NE
    region). If we considered c2 as an interstellar cloud, cloud
    evaporation might play an important role in its high density.
    Based on the cloud evaporation model, however, high $kT_{\rm e}$
    components are considered as the emission from the postincident
    shocked region just as the same case of the previous model. Thus, we
    met with a similar difficulty in trying to explain the X-ray 
    properties by the cloud
    evaporation model.

    We next calculated the cooling time of the posttransmitted shocked
    gas.  For clouds with density greater than $n_c$, the
    posttransmitted shocked gas rapidly cools down to $10^5$ K, and
    there is no X-ray emission. Based on the equation
    (\ref{eq:review:evapo:criticalDensity}), we calculated the maximum
    cloud thickness, $R_4$, to be heated by the transmission shock to
    emit X-rays.  Figure~\ref{fig:dis:east:nc} shows the maximum cloud
    radius plotted as a function of $\alpha$ and $n_0$. Vertical dashed
    lines show the permitted area for $\alpha$ (inferred in the
    subsequent subsection) while horizontal solid lines show the
    permitted area for $n_0$. This figure shows that the transmission
    shock wave will cool down to below the X-ray emitting temperature
    when it moves into a cloud roughly 0.02$-$3 pc in size.  This
    indicates that the transmission shock can heat up roughly half of c2
    for which the radius is roughly 2.7 pc and it cannot pass through c2
    since it will cool down to below the X-ray emitting temperature.
    Therefore, the cloud evaporation model can be ruled out as an
    explanation for the hard X-ray emitting gas, although it might
    account for the low-$kT_{\rm e}$ component for future observations.

    \subsubsection{Reflection and Transmission Shock Model} 

    As described in the previous section, three kinds of shock waves might
    occur when the blast wave interacts with the cloud. We suppose that
    the posttransmitted and the postreflected shocked regions correspond
    to the soft and hard X-ray regions, respectively, in case of the
    eastern region. The reflection shock propagates into the
    postincident shocked region and provides an additional heating.
    The transmission shock
    propagates into the high density cloud and heats it up to X-ray
    emitting temperatures if the density contrast between the cloud and
    the ambient matter is not too high, accounting for the bright X-rays 
    seen in the ROSAT band.

   Based on this model, the ratio of $kT_{\rm e}$ for the postreflected
   shocked region to that for the posttransmitted shocked region depends
   on the density contrast.  Based on Figure \ref{fig:dis:east:sgro} we
   calculated the ratio for $kT_{\rm e}$ and $n_e$ between the
   postreflected shocked region and the posttransmitted shocked region,
   shown in Figure~\ref{fig:dis:east:ratio}.  We also plotted our
   results in Figure~\ref{fig:dis:east:ratio} for c1 and r1$-$r4. The
   density contrast, $\alpha$, can be determined based both on the
   $kT_{\rm e}$ ratios and the $n_e$ ratios as shown in
   Table~\ref{table:dis:east:densityContrast}. We can obtain a
   consistent value of $\alpha \simeq 10$ for all regions even though
   the statistical uncertainties are large for the $n_e$
   ratios. Therefore, based on the reflection shock model, we can
   quantitatively explain the observed values of $kT_{\rm e}$ for high
   and low $kT_{\rm e}$ components with an interstellar cloud for which
   the density is roughly an order of magnitude larger than that of the
   ambient medium.

   We can calculate the various physical parameters of the postreflected
   and posttransmitted shocked region based on
   Figure~\ref{fig:dis:east:ratio}.
   Table~\ref{table:dis:east:densityContrast} shows values of $n_2 /
   n_1$ (= $\alpha _{\rm r}$ ) and $T_2 / T_1$ for each region. Based on
   these results, $n_e$ and $kT_{\rm e}$ at the postreflected
   shocked region is roughly 1.8 and 1.5 times larger than those at the
   postincident shocked region.  Therefore, $n_{\rm e}$ and $kT_{\rm e}$
   at the postincident shocked region are $\sim$ 2.8 cm$^{-3}$ and
   $\sim$ 0.21~keV. This implies that $n_0$ at the eastern region is 0.7
   cm$^{-3}$. This value is roughly three times larger than the mean
   value (Ku et al. 1984). This might be a region of enhanced density,
   a `tail' of the cloud, extending around c2 as suggested by HC.

   If we compare this density with that obtained at the NE region,
   $kT_{\rm e}$ at the eastern region is found to be
   $(0.7/0.25)^{-2/5} \times 0.3\ \approx$ 0.20~keV.  This value is in
   good agreement with our result.

   Thus, we conclude that the reflection shock and transmission shock
   can explain the X-ray properties observed both with ROSAT and with
   ASCA.

   It is interesting to note that the density contrast obtained at the
   eastern region is consistent with that obtained at the NE
   region (MT). Combining these results, we suppose the existence of a
   large-scale high density region from the NE toward the
   eastern region of the Cygnus Loop.

  \section{SHOCK STRUCTURE OF THE NORTHERN REGION}

   \subsection{Complicated X-ray Surface Brightness and $kT_{\rm e}$ Map}

   The X-ray image of the \north\ shows a peculiar V-shaped feature, as
   shown in Figure~\ref{fig:ana:limb:image}.  This V-shaped region is
   the brightest among the regions with the same radius from the center.
   The $kT_{\rm e}$ distribution of the \north\ is not uniform as shown
   in Figure~\ref{fig:dis:north:kt}. We found that three regions showed
   $kT_{\rm e}$ (0.5$-$0.6 keV) higher than other regions (0.3$-$0.4
   keV). We refer to these three regions as h1$-$h3 as marked in
   Figure~\ref{fig:dis:north:kt} with green lines.  Comparing the
   surface brightness map with $kT_{\rm e}$ map, we noticed a strong
   anti-correlation between them.

  \subsection{Emission Mechanism of the Northern Region}

  We centered our FOV on the brightest part in the northern region
  which was about 20\arcmin\ inside the shock front.  As mentioned in
  section~\ref{sec:dis:east:em}, we expected that the brightest region
  was roughly 4\arcmin\ inside the shock front if we assumed the blast
  wave to be propagating into a homogeneous medium. The apparent
  structure of the \north\ is unlikely to be explained by such a simple
  blast wave model.

  As mentioned in the previous section, the \north\ is the brightest
  among the regions with the same radius from the center. One of the
  following two scenarios may explain the emission
  mechanism; the high density medium may be like an interstellar cloud
  (evaporation or reflection shock heating) or the shock front structure
  may be due to a projection effect. In the subsequent section, we will
  investigate the possible explanation for the current X-ray features
  observed both with ROSAT and with ASCA.

   \subsection{Shock Front Structures due to the Projection Effect}

   The mean $kT_{\rm e}$ around the V-shaped region is roughly
   0.4~keV. It is similar to that obtained at the NE region (MT).
   However, the surface brightness is much higher than regions with the
   same radius in the direction of the NE region.  If the V-shaped
   region comes from the shock front due to a projection effect, the
   mean density of the \north\ must be higher assuming the pressure
   equilibrium. Therefore, we can expect a lower $kT_{\rm e}$ than that
   of the NE region assuming the pressure equilibrium. However, we found
   $kT_{\rm e}$ to be similar to each other. We conclude that the bright
   X-ray appearance does not come from a projection effect of the shock
   front structures.

   \subsection{Cloud Evaporation Model}

   We next investigate a cloud evaporation model. After the interstellar
   cloud is engulfed by the shock wave, it begins to evaporate when the
   cloud density is higher than the critical density for direct shock
   heating. The value of $kT_{\rm e}$ of the shocked cloud ($kT_3$)
   depends on the density contrast between the cloud and the ISM. The
   value of $kT_3$ must be lower than that of the postincident shocked
   region.  Assuming that the cloud is sitting at the center of our FOV,
   we can expect bright X-ray emission from the cloud. This picture
   resembles our results at the point of 1) the bright and low $kT_{\rm
   e}$ emission at the center of our FOV and 2) high $kT_{\rm e}$
   emission at h2 (and possibly h3). Generally speaking, the value of
   $kT_{\rm e}$ decreases toward the shock front where the density is
   the highest. Therefore, h1 is considered as the postincident shocked
   region shocked by the blast wave passing over the cloud.  We then
   investigated whether or not we could explain the X-ray feature of h1
   by a model that included direct shock heating.

   The angular distance from the center of the Cygnus Loop to h1 is
   roughly 72\arcmin.  The value of $kT_{\rm e}$ obtained at the region
   with the same radius in the direction of the NE region is roughly
   0.4~keV (MT). If such a high $kT_{\rm e}$ medium at h1 is the shock
   heated ISM, we can estimate the density by using the Sedov
   model. Since $kT_{\rm e}$ of the shock wave is proportional to
   $n^{-2/5}$, $n_0$ of the \north\ is expected to be
   $(0.6/0.4)^{-5/2}\times 0.25 \approx 0.09$ cm$^{-3}$.  The surface
   brightness of h1 is lower by a factor of $\sim$ 3 than that of the NE
   region, which is consistent with the low density derived
   above. However, based on a similarity solution, the shock radius is
   proportional to $n^{-1/5}$. The current location of the shock front
   passing over the cloud is expected to be (0.25/0.07)$^{1/5}\approx$
   1.3 farther away from the center than that at the NE region. However,
   it is not the case as shown in Figure~\ref{fig:ana:limb:image}.
   Therefore, it seems difficult to explain the current X-ray feature of
   h1 with the model of shock heating by the incident shock wave.

   Therefore, even if we take into account not only a single cloud but
   also multiple clouds, we expect that the cloud evaporation model would
   be ruled out to explain the \north.

   \subsection{Reflection Shock Model}

   We found that neither the direct shock heating model nor the cloud 
   evaporation model could account for the \north\  of the Cygnus Loop. We 
   next consider the reflection shock heating model.

   There are several configurations for the location of the cloud. Among
   them, we choose two favorable configurations as discussed in the
   subsequent sections.

    \subsubsection{Cloud at the Northern Region in Our FOV (h1)} 

   We first examine the case that a cloud exists at h1. We can interpret
   the structure of the V-shaped region as the postreflected shocked
   region as shown in Figure~\ref{fig:review:reflection:geometry}.
   Regions of h2 and h3 are expected to be the postincident shocked
   region. In this case, we expect $kT_{\rm e}$ at the V-shaped region
   to be higher than those of h2 and h3. However, this is not the
   case. Therefore, this is unlikely to be explained with the reflection
   shock model.

   The value of $kT_{\rm e}$ of h1 is much higher than those at other
   regions at the same angular distance from the center. Such high
   $kT_{\rm e}$ is expected in the postreflected shocked region. If a
   cloud is at h1, we will expect the emission from the cloud itself at
   h1 either in at soft X-ray or optical wavelengths.

   The value of $kT_{\rm e}$ of the posttransmitted shocked region
   depends on the density contrast. For a small density contrast, we
   expect that a low $kT_{\rm e}$ component is bright in the soft X-ray
   energy band (ROSAT band). On the contrary, for the case of a high
   density contrast, $kT_{\rm e}$ of the posttransmitted shocked region
   decreases so that there is no X-ray emission, so that we then expect
   some optical counterparts to be seen at the position of the
   cloud. Based on the $kT_{\rm e}$ map obtained with the ROSAT all-sky
   survey (Aschenbach \& Leahy 1994), $kT_{\rm e}$ at h1 region is
   higher than those at its neighboring regions. This will deny the
   existence of the soft X-ray emitting component.

   Figure~\ref{fig:dis:north:optical} shows the optical image of the
   \north\ of the Cygnus Loop which is a digital sky survey image
   retrieved from the {\tt Skyview} program, supported by HEASARC/GSFC.
   There are some optical counterparts seen. However, we find no
   emission around the h1 region. Therefore, we conclude that there is
   not a low $kT_{\rm e}$ component at regions neighboring h1 either in
   the soft X-ray region or at optical wavelengths.

   Thus, this configuration cannot explain either the low $kT_{\rm e}$ of
   the V-shaped region or the lack of low $kT_{\rm e}$ component at h1.

    \subsubsection{Cloud at the Center of Our FOV} 

   We next consider a cloud at the center of our FOV.  After the
   interaction between the blast wave and the cloud, the blast wave
   passes over the cloud. The reflection shock propagates into the
   postincident shocked region while the transmission shock propagates
   into the cloud.  The current geometry we suppose for the \north\ is
   schematically shown in Figure~\ref{fig:dis:north:geometry}.

   In this configuration, we directly observe the postreflected shocked
   region with a higher $kT_{\rm e}$ among the neighboring mediums at
   two different regions. Between these two regions, we observe the
   overlapping of a postreflected shocked region and a posttransmitted
   shocked region. Because the density of both regions is high, we
   expect substantially bright X-ray emission at the overlapping
   region. The value of $kT_{\rm e}$ at the overlapping region is
   intermediate between the low $kT_{\rm e}$ (k$T_3$) the high $kT_{\rm
   e}$ ($kT_2$).

   If we apply this configuration to our data, we suppose that h1$-$h3
   come from the postreflected shocked region as marked by upper and
   lower broken lines. Between h1 and h2, the bright V-shaped region can
   be interpreted as the overlapping region.  Therefore, we favor this
   picture since this picture can explain both the high X-ray surface
   brightness seen at the center of the \north\ obtained with ROSAT and
   the $kT_{\rm e}$ distribution determined with the ASCA SIS.

   We extracted spectra at h1$-$h3 regions and fitted them with a single
   component NEI model.  However, we failed to fit them with a single
   component model. As shown in Figure 7 there are still spectral
   variations in each region, possibly because there are some amount of
   contamination from the low-$kT_{\rm e}$ component and also the model
   we proposed is too much simplified. However, we believe the overall
   plasma structure can be explained with the current configuration.

   We estimate other physical parameters based on this model assuming a
   spherical cloud. The separation between h1 and h2 roughly corresponds
   to the diameter of the cloud. The cloud is then at ($\alpha=20^{\rm
   h} 49^{\rm m} 57^{\rm s}, \delta =32^\circ 3^\prime
   33^{\prime\prime}$) with a radius of roughly 7$^\prime \simeq$
   1.6~pc. The values obtained for $kT_{\rm e}$ at h1$-$h3 are
   considered as $T_2$ whereas that of V-shaped region ($\sim$ 0.4 keV)
   stands for the mean value of $T_2$ and $T_3$. Therefore, we extracted
   the X-ray spectrum at the V-shaped region and fitted it with a two
   component NEI model with different $kT_{\rm e}$. The values
   obtained for $kT_{\rm e}$ are $\sim$ 0.2 and $\sim$ 0.5 keV. The
   value of the high $kT_{\rm e}$ component agrees reasonably with that
   obtained at h1$-$h3. Thus, $T_3$ can be considered as $\simeq$ 0.2
   keV.

   The ratio of $T_2$ to $T_3$ depends on the density contrast as
   discussed in section 4. Based on Figure~\ref{fig:dis:east:ratio}, we
   obtain the density contrast $\alpha$ to be $\approx$ 5.  The cloud
   density is estimated to be roughly 1 cm$^{-3}$ assuming the ambient
   density of 0.2 cm$^{-3}$. For $\alpha=5$, we obtain $kT_{\rm e}$ of
   postincident shocked region $T_1$ of 0.45~keV and $\beta$=2. $T_1$
   obtained is similar to that of the region with the same angular
   distance from the center to the direction of the NE region.

   Figure~\ref{fig:dis:north:radius} shows the maximum radius of the
   cloud which can be heated up directly by the shock wave, in the
   $\alpha- T_1$ plane.  The vertical solid line represents the value of
   $\alpha$ at the \north\ region. As shown in this figure, the cloud
   with size of $\leq$ 1.6 pc will be destroyed by the blast wave with
   $kT_1 \geq 0.2$ keV. On the contrary, the value of $kT_1$ we obtained
   is 0.45 keV.  Such a cloud cannot survive during passage of the blast
   wave. Therefore, we can rule out the cloud evaporation model to
   explain the bright X-ray features seen at \north.

   In this way, the reflection shock model can well explain the
   observed X-ray features with a simple geometry as shown in
   Figure~\ref{fig:dis:north:geometry}.

   \section{CONCLUSION}

   We have performed high quality X-ray observations of the Cygnus
   Loop using the ASCA Observatory and the ROSAT Observatory. We have
   found that the X-ray surface brightness and $kT_{\rm e}$ significantly
   vary inside our FOV. They cannot be explained with a simple blast
   wave model. We carried out calculations for the interaction between
   the blast wave and the interstellar cloud in order to determine the
   characteristics of the postincident shocked gas, the posttransmitted
   shocked gas, and the reflection shocked gas. We applied the
   evaporation model as well as the reflection (and transmission) shock
   model for the eastern and the northern regions. We have difficulties
   explaining the X-ray nature with the evaporation model for both
   regions. Only the reflection shock model can quantitatively explain
   the observed $kT_{\rm e}$ and the X-ray surface brightness.

   The cloud size is difficult to estimate. However, it is at least 1 pc
   in size for both regions. Such a large cloud is sitting in the
   NE region as well (MT). Combining all of these results, we conclude
   that a cloud of
   large scale is covering at least one third of the Cygnus Loop.
   This fact strongly supports the existence of a pre-existing cavity
   wall, possibly created by the precursor.

   \acknowledgments

   We would like to thank the referee Dr. David Burrows for many useful
   comments and suggestions for improving the paper.
   We are grateful to all the other members of the ASCA team.
   Dr. B. Aschenbach kindly supplied us the whole X-ray image of the
   Cygnus Loop obtained with the ROSAT all-sky survey.  We acknowledge
   to Mr. C. Baluta for his critical reading of the manuscript.  This
   research is partially supported by ACT-JST Program, Japan Science and
   Technology Corporation.  Part of this research has made use of data
   obtained through the High Energy Astrophysics Science Archive
   Research Center Online Service, provided by the NASA/Goddard Space
   Flight Center.

\clearpage

\begin{deluxetable}{cccccccc}
\tabletypesize{\scriptsize}
\tablecaption{\sc Observational Log of the Cygnus Loop \label{obslog}}
\tablewidth{0pt}
\tablehead{
\colhead{Date} & \colhead{Sequence Number} & \colhead{Coordinate (J2000)} &
\colhead{Position} & \colhead{CCD mode$^a$}  & \colhead{Exposure (ks)} &
\colhead{Intensity (c s$^{-1}$)} 
}
\startdata
  1993 Oct 18\dotfill & 50008010 & 20$^{\rm h}$50$^{\rm m}$ 2.$^{\!\rm s}$4,
 32$^{\rm d}$ 3$^{\prime}$ 4$^{\prime\prime}$ &
  pos-NN & 0123 / 2301 & 5.5 / 6.5 & 12.44 / 12.13 \\
  1994 Nov 01\dotfill & 52003020 & 20$^{\rm h}$57$^{\rm m}$ 7.$^{\!\rm s}$2,
  31$^{\rm d}$ 14$^{\prime}$ 56$^{\prime\prime}$ &
  pos-EU & 0123 / 2301 & 3.7 / 5.1 & 6.60 / 5.78  \\
  1994 Nov 01--02\dotfill & 52003050 & 20$^{\rm h}$57$^{\rm m}$
 22.$^{\!\rm s}$3,  30$^{\rm d}$ 53$^{\prime}$ 13$^{\prime\prime}$ &
  pos-ED & 0123 / 2301 & 5.9 / 6.5 & 3.02 / 2.73 \\
  1994 Nov 02\dotfill & 52003000 & 20$^{\rm h}$55$^{\rm m}$ 30.$^{\!\rm s}$0,
  30$^{\rm d}$ 56$^{\prime}$ 49$^{\prime\prime}$ &
  pos-ER & 0123 / 2301 & 2.9 / 4.3 & 5.16 / 4.00 \\
 \enddata

\tablenotetext{a}{CCD mode shows the CCD ID number used in the
                observation.}

\tablecomments{The columns which have two values divided by `/'
        show the value of SIS-0 and SIS-1, respectively.}

\end{deluxetable}

\clearpage

\begin{deluxetable}{rccccccc}
 \tabletypesize{\scriptsize}
 \tablecaption{\sc Spectral Features at the Eastern Region with the Combined
 Spectral Analysis of the ASCA SIS and the ROSAT PSPC \label{east_fit_result}}
 \tablewidth{0pt}
 \tablehead{
 \colhead{Region} & \colhead{$kT_{\rm e}$-1}   & \colhead{log($\tau$)}  &
 \colhead{EM-1} &  \colhead{$kT_{\rm e}$-2}  & \colhead{EM-2} &
 \colhead{$N_{\rm H}$} & \colhead{reduced $\chi^2$ (dof)} \\
 \colhead{}   & \colhead{(keV)} &\colhead{} &\colhead{(${\rm cm^{-6}\ pc}$)} &
 \colhead{(eV)} & \colhead{(${\rm cm^{-6}\ pc}$)} &
 \colhead{($10^{20}$ cm$^{-2}$)} & \colhead{}}
 \startdata
 c1\dotfill & \errorBoth{0.27}{0.07}{0.08} &
 \errorBoth{10.9}{2.1}{0.4} &
 \errorBoth{20}{90}{10} &
 \errorBoth{43}{3}{1} &
 (\errorBoth{1.0}{0.7}{0.5})$\times 10^4$ &
 \error{8.8}{0.7} & 1.2 (116) \\
 r1\dotfill & \error{0.34}{0.04} &
 \error{11.4}{0.3} &
 \errorBoth{9}{8}{5} & 
 \errorBoth{54.1}{0.4}{0.6} &
 (\errorBoth{3.0}{0.6}{0.2})$\times 10^3$ &
 \errorBoth{8.6}{0.4}{0.5} & 1.2 (144) \\
 r2\dotfill & \errorBoth{0.34}{0.04}{0.03} &
 \errorBoth{11.9}{1.1}{0.3} &
 \errorBoth{3}{7}{2} & \errorBoth{67}{5}{13} &
 (\errorBoth{4}{10}{2})$\times 10^2$ &
 \errorBoth{7.3}{1.5}{0.9} & 1.3 (143) \\
 r3\dotfill & \errorBoth{0.31}{0.05}{0.09} &
 \errorBoth{11.7}{0.5}{0.3} &
 \errorBoth{2}{6}{1} &
 \errorBoth{54}{4}{20} &
 (\errorBoth{3}{10}{2})$\times 10^3$ &
 \errorBoth{8.7}{1.1}{0.8} & 1.3 (141) \\
 r4\dotfill & \errorBoth{0.32}{0.06}{0.05} &
 \errorBoth{10.8}{1.1}{0.2} &
 \errorBoth{12.3}{10}{0.1} &
 \errorBoth{47}{10}{4} &
 (\errorBoth{4}{8}{3})$\times 10^3$ &
 \errorBoth{9}{1}{2} & 1.1 (121) \\
 \enddata

\end{deluxetable}

\clearpage

\begin{deluxetable}{rccccc}
 \tabletypesize{\normalsize}
 \tablecaption{\sc Abundances of Heavy Elements Obtained at the Eastern
 Region \label{east_abundance}}
 \tablewidth{0pt}
 \tablehead{
 \colhead{Region} & \colhead{C, N, O}   & \colhead{Ne}  &
 \colhead{Mg} &  \colhead{Si}  & \colhead{Fe}}
 \startdata
   c1\dotfill & \errorBoth{0.2}{0.3}{0.1} &  \errorBoth{0.3}{0.4}{0.2} &
 \errorBoth{0.25}{0.5}{0.23} &  $<10$ &  \errorBoth{0.2}{0.4}{0.1} \\
   r1\dotfill &  \errorBoth{0.8}{0.3}{0.2} &  \errorBoth{1.7}{0.9}{0.3} &
 \errorBoth{1.0}{0.5}{0.3} &  \errorBoth{0.9}{0.9}{0.6} &
 \errorBoth{0.50}{0.22}{0.04} \\
   r2\dotfill &  \errorBoth{0.8}{3.8}{0.5} &  \errorBoth{1.0}{6.2}{0.6} &
 \errorBoth{0.9}{3.4}{0.6} &  \errorBoth{1.3}{5.4}{0.8} &
 \errorBoth{0.5}{1.8}{0.2} \\
   r3\dotfill &  \errorBoth{6}{1}{5} &  $<$ 10 &  \errorBoth{6}{2}{5} &
   $<$ 6 &  \errorBoth{2.8}{0.2}{2.3} \\
   r4\dotfill &  \errorBoth{0.26}{0.12}{0.08} &
 \errorBoth{0.3}{0.2}{0.1} & $<$ 0.3 &  $<$ 1 &
 \errorBoth{0.19}{0.08}{0.05} \\
 \enddata

\end{deluxetable}

\clearpage

\begin{deluxetable}{rcccccc}
 \tabletypesize{\small}
 \tablecaption{\sc Volume, Density, and Pressure of the Eastern Region \label{table:dis:east:pspcSpec}}
 \tablewidth{0pt}
 \tablehead{
 \colhead{Region} & \multicolumn{2}{c}{High $kT_{\rm e}$}   & \colhead{}  &
 \multicolumn{2}{c}{Low $kT_{\rm e}$} &  \colhead{Pressure (p/k)}  \\
 \cline{2-3}\cline{5-6}
 \colhead{} & \colhead{Density} &  \colhead{Volume} &  \colhead{} &
 \colhead{Density} &  \colhead{Volume} &  \colhead{} \\
 \colhead{} & \colhead{(cm$^{-3}$)} &  \colhead{(pc$^{3}$)} &  \colhead{} &
 \colhead{(cm$^{-3}$)} &  \colhead{(pc$^{3}$)} &
 \colhead{($10^7$ cm$^{-3}$ K)}
 }
 \startdata
	c1 & \errorBoth{6}{3}{2} & 0.50 & &
		\error{40}{10} & 7.4 &
                \error{4}{1}\\
	r1 & \error{4.6}{0.6}  & 0.45 & &
                \errorBoth{29}{3}{1} & 3.7   &\errorBoth{3.6}{0.3}{0.1} \\
	r2 &  \errorBoth{1.9}{2.2}{0.7}  & 0.79 & &
                \errorBoth{10}{11}{3} & 3.7 & \errorBoth{1.5}{1.7}{0.5} \\
	r3 & \errorBoth{5}{5}{3} & 0.073 &  &
		 \errorBoth{30}{30}{9} & 3.7 &\errorBoth{4}{4}{2} \\
	r4 &  \errorBoth{5}{4}{3} & 0.56 & &
		\errorBoth{30}{30}{20} & 3.7 & \errorBoth{3}{3}{2} \\
 \enddata

\end{deluxetable}

\clearpage

\begin{deluxetable}{rcccc}
 \tabletypesize{\small}
 \tablecaption{\sc Characteristics of the Reflection and the Transmission
 Shocked Gas at the Eastern Region \label{table:dis:east:densityContrast}}
 \tablewidth{0pt}
 \tablehead{
 \colhead{Region} & \colhead{$\alpha$ (based on $kT_{\rm e}$ ratio)} &
 \colhead{$\alpha$ (based on $n_e$ ratio)} &
 \colhead{$n_2 / n_1$ ($\alpha _r$) $^a$} &  \colhead{$T_2 / T_1\ ^a$}
  }
 \startdata
	c1 & \error{11}{4} & \errorBoth{11}{3}{1} &
             \errorBoth{1.8}{0.1}{0.2} & \error{1.5}{0.1} \\
	r1 & \errorBoth{11}{2}{1} & \errorBoth{11}{15}{4} &
		\error{1.8}{0.1} & \errorBoth{1.54}{0.04}{0.02} \\
	r2 & \errorBoth{9}{2}{1} & $>3$ & \error{1.7}{0.1} & 
		\error{1.5}{0.1} \\
	r3 & \error{10}{5} & $>4$& \errorBoth{1.7}{0.1}{0.2} & 
		\error{1.5}{0.1} \\
	r4 & \errorBoth{12}{3}{4} & $>5$ &  \error{1.8}{0.1} & 
		\error{1.6}{0.1} \\
 \enddata

\tablenotetext{a}{These values are calculated using $\alpha$ extracted
 from $kT_{\rm e}$ ratio.}

\end{deluxetable}

\clearpage

\begin{figure}

\centerline{FIGURE CAPTIONS}

 \caption{X-ray surface brightness map of the Cygnus Loop
 obtained during the ROSAT all-sky survey (Aschenbach \& Leahy 1994). The
 black squares show the FOVs of the eastern and northern portion
 as observed with ASCA. } \label{fig:ana:limb:image}

 \caption{X-ray spectra obtained at pos-NN (a), pos-EU (b), pos-ED (c),
 pos-ER (d). X-ray emission is dominated by thermal emission and emission
 lines from Mg and Si are clearly resolved.  Line identifications are
 also shown in (a).} 
 \label{all_spec}

 \caption{(a) $kT_{\rm e}$ and (b) log($\tau$) maps of the eastern region
 obtained with the SIS.  North is up and east is left. } 
 \label{fig:ana:east:kt}
 
 \caption{X-ray surface brightness map of the eastern region obtained
 with the ROSAT PSPC (upper) and the ASCA SIS (lower).  Black squares
 show the FOVs of the SISs and a black circle marks the bright clump
 (c1).  The difference between the surface brightness in the PSPC and
 that in the SIS energy bands is clearly seen which is due to the
 difference of the energy bands.}  \label{fig:ana:east:pspc}

 \caption{Band ratio map of the eastern region of the Cygnus
 Loop. This map is extracted as (0.44$-$2.0~keV) / (0.11$-$0.28~keV).
 The region of c1 and the soft X-ray emitting region (c2) are encircled
 by small and large circles, respectively. The regions where spectra
 (r1$-$r4) are extracted are shown.}  \label{fig:ana:east:hardness}

 \caption{ASCA SIS and ROSAT PSPC spectra at c1 (a) and r1$-$r4 (b--e)
 regions.  Best fit curves are shown with solid lines.}
 \label{fig:ana:east:pspcSpec}

 \caption{Same as Figure \ref{fig:ana:east:kt} but of the \north.  Three
 high $kT_{\rm e}$ regions are marked by green circles, named h1$-$h3.}
 \label{fig:dis:north:kt}

\end{figure}

\begin{figure}

 \caption{X-ray spectrum obtained from the portion of the remnant
 observed with the FPCS. Each component is also shown by a solid curve.}
 \label{fig:ana:sis:north:bg}

  \caption{Geometry of the reflection shock. Upper panel shows the
  cross section of the interaction with the cloud. Lower panel shows
  the density structure along the cross section of the cloud.}
  \label{fig:review:reflection:geometry}

  \caption{Physical parameters for the reflection shock as a function of
  the density contrast between the cloud and the interstellar medium.}
  \label{fig:dis:east:sgro}

  \caption{Maximum cloud thickness which can be heated up by the blast
  wave is plotted versus $\alpha$ and $n_0$.  The value of $T_1$ is
  assumed to 0.21 keV.  Contours are shown in a logarithmically scaled
  from $1\times10^{-4}$ to $1 \times 10^4$ pc. The cloud size of c2 (2.7
  pc) is also shown.  Vertical dashed lines show the permitted area for
  $\alpha$ in the eastern region while horizontal solid lines show the
  permitted area for $n_0$.}  \label{fig:dis:east:nc}

  \caption{Ratios of the temperatures and of the densities between the
  postreflected and posttransmitted shocked regions.  We should note
  that the density ratios plotted for r2, r3, and r4 only show the upper
  limits.}\label{fig:dis:east:ratio}

  \caption{Digital sky survey image of the \north\ of the Cygnus
  Loop obtained with the {\tt Sky View}. A black square shows the FOV of
  the SIS.}  \label{fig:dis:north:optical}

  \caption{Geometry of the interaction between the blast wave and the
  interstellar cloud which accounts for the X-ray appearance of the
  \north.} \label{fig:dis:north:geometry}

  \caption{Same as Figure \ref{fig:dis:east:nc} but the vertical axis
  shows $T_1$. Ambient gas density is fixed to 0.2 cm$^{-3}$.  Contours
  are shown in a logarithmically scaled and the cloud size of 1.6 pc is
  also shown. Vertical solid line represents the value of $\alpha$
  obtained at the \north.}  \label{fig:dis:north:radius}

\end{figure}


\begin{thebibliography}{}
    \bibitem{} Aschenbach, B., \& Leahy, D.A. 1999, \aap, 341, 602
    \bibitem{} Ballet, J., Arnaud, M., \& Rothenflug, R. 1984, \aap,
	    133, 357
    \bibitem{} Braun, R., \& Strom, R.G. 1986, \aap, 164, 208
    \bibitem{} Charles, P.A., Kahn, S.M., \& McKee, C.F. 1985, \apj, 295, 456
    \bibitem{} Cox, D.P. 1972, \apj, 178, 169
    \bibitem{} Decourchelle, A., Sauvageot, J.L., Ballet, J., \& Aschenbach,
	    B. 1997, \aap, 326, 811
    \bibitem{} Dotani, T. et al. 1995, ASCA Letter News 3, 25
    \bibitem{} Dotani, T. et al. 1997, ASCA Letter News 5, 14
    \bibitem{} Fesen, R.A., Blair, W.P., \& Kirshner, R.P. 1982, \apj,
	    262, 171
    \bibitem{} Gehrels, N., \& Williams, E.D. 1993, \apj, 418, L25
    \bibitem{} Graham, J.R. Levenson, N.A., Hester, J.J., Raymond,
	    J.C. \& Petre, R. 1995, \apj, 444, 787
    \bibitem{} Hester, J.J., \& Cox, D.P. 1986, \apj, 300, 675 (HC)
    \bibitem{} Hester, J.J., Raymond, J.C., \& Blair, W.P. 1994, \apj,
	    420, 721
    \bibitem{} Inoue, H., Koyama, K., Matsuoka, M., Ohashi, T., Tanaka,
	    Y., \& Tsunemi, H. 1979, in X-Ray Astronomy,
	    ed. W.A. Baity \& L.E. Peterson (Pergamon Press, Oxford), 309
    \bibitem{} Levenson, N.A. et al., 1997, \apj, 484, 304
    \bibitem{} Kahn, S.M., Charles, P.A., Bowyer, S., \& Blissett, R.J. 1980,
	    \apjl, 242, L19
    \bibitem{} Klein, R.I., McKee, C.F., \& Colella P. 1994, \apj, 420, 213
    \bibitem{} Ku, W.H.-M., Kahn, S.M., Pisarski, R., \& Long, K.S. 1984,
	    \apj, 278, 615
    \bibitem{} Masai, K. 1984, \apss, 98, 367
    \bibitem{} Masai, K. 1994, \apj, 437, 770
    \bibitem{} McCray R., Snow T.P.Jr. 1979, \araa, 17, 213
    \bibitem{} McKee, C.F., \& Cowie L.L. 1975, \apj, 195, 715
    \bibitem{} McKee, C.F., \& Ostriker, J.P. 1977, \apj, 218, 148
    \bibitem{} McKee, C.F. 1987, in Supernova Remnants and the Interstellar
	    Medium, p205 
    \bibitem{} Miyata, E., Tsunemi, H., Pisarski, R., \& Kissel, S.E. 1994, 
	    \pasj L, 46, L101
    \bibitem{} Miyata, E., Tsunemi, H., Kohmura, T., Suzuki, S., \& Kumagai, S.
	    1998, \pasj, 50, 2
    \bibitem{} Miyata, E., \& Tsunemi, H. 1999, \apj, 525, 305 (MT)
    \bibitem{} Raymond, J.C., Hester, J.J., Cox, D., Blair, W.P., Fesen,
	    R.A., \& Gull, T.R. 1988, \apj, 324, 869
    \bibitem{} Sedov, L.I. 1959, Similarity and Dimensional
	    Methods in Mechanics, 10th ed.
	    (New York: Academic Press)
    \bibitem{} Serlemitsos, P.J., Jalota, L., Soong, Y., Kunieda, H.,
	    Tawara, Y., Tsusaka, Y., Suzuki, H., Sakima, Y.
	    et al. 1995, \pasj, 47, 105
    \bibitem{} Sgro, A.G. 1975, \apj, 197, 621
    \bibitem{} Shull, P.Jr., \& Hippelein, H. 1991, \apj, 383, 714
    \bibitem{} Spitzer, L. 1962, Physics of Fully Ionizaed Gases
	    (New York Interscience)
    \bibitem{} Spitzer, L. 1982, \apj, 262, 315
    \bibitem{} Szentgyorgyi, A.H., Raymond, J.C., Hester, J.J., \&
	    Curiel, S. 2000, \apj, 529, 279
    \bibitem{} Tanaka, Y., Inoue, H., \& Holt, S.S. 1994, \pasj, 46, L37
    \bibitem{} Teske, R.G. 1990, \apj, 365, 256
    \bibitem{} Teske, R.G., \& Kirshner, R.P. 1985, \apj, 292, 22
    \bibitem{} Tsunemi, H., \& Inoue, H. 1980, \pasj, 32, 247
    \bibitem{} Vedder, P.W., Canizares, C.R., Markert, T.H., \& Pradhan, A.K.
         1986 \apj, 307, 269
    \bibitem{} Yamashita, A. Dotani, T., Bautz, M.W., Crew, G., Ezuka, H.,
	    Gendreau, K., Kotani, T., Mitsuda, K., Otani, C., Rasmussen, A.,
	    Ricker, G., \& H. Tsunemi, 1997, IEEE Trans. NS, 44, 847
   \end{thebibliography}
\end{document}